\documentclass[preprint]{aastex}
\usepackage{emulateapj5}
\usepackage{natbib}

\input psfig.tex
\def\aa{{ A\&A}}

\def\aj{{AJ}}
\def\al{$\alpha$}

\def\annrev{{ARA\&A}}
\def\apj{{ApJ}}
\def\apjs{{ApJS}}
\def\asec{$^{\prime\prime}$}

\def\deg{$^{\circ}$}

\def\e#1{ $\times$ 10$^{#1}$}

\def\kms    {~km~s$^{-1}$}

\def\lum{erg s$^{-1}$}

\def\mnras{{MNRAS}}

\def\pasp{{PASP}}
\def\percm2{cm$^{-2}$}

\def\lax    {${_<\atop^{\sim}}$ }
\def\gax    {${_>\atop^{\sim}}$ }

\def\arcsec{\hbox{$^{\prime\prime}$}}

\def\farcs{\hbox{$.\!\!^{\prime\prime}$}}

\def\hii{\ion{H}{2}}

\slugcomment{To appear in The Astrophysical Journal (Letters).}
\shorttitle{{\it Chandra} Survey of Nearby Galaxies}
\shortauthors{HO ET AL.}

\begin{document}

\title{Detection of Nuclear X-ray Sources in Nearby Galaxies with 
{\it Chandra}}

\author{
Luis~C.~Ho\altaffilmark{1}, 
Eric~D.~Feigelson\altaffilmark{2}, 
Leisa~K.~Townsley\altaffilmark{2}, 
Rita~M.~Sambruna\altaffilmark{2,3},
Gordon~P.~Garmire\altaffilmark{2}, 
W.~N.~Brandt\altaffilmark{2},
Alexei~V.~Filippenko\altaffilmark{4}, 
Richard~E.~Griffiths\altaffilmark{5}, 
Andrew~F.~Ptak\altaffilmark{5}, and
Wallace~L.~W.~Sargent\altaffilmark{6} 
}

\altaffiltext{1}{The Observatories of the Carnegie Institution of Washington, 
813 Santa Barbara St., Pasadena, CA 91101-1292.}

\altaffiltext{2}{Department of Astronomy and Astrophysics, The
Pennsylvania State University, 525 Davey Lab, University Park, PA 16802.}

\altaffiltext{3}{George Mason University, Department of Physics and Astronomy, 
MS 5C3, Fairfax, VA 22030-4444.}

\altaffiltext{4}{Department of Astronomy, University of California, Berkeley, 
CA 94720-3411.}

\altaffiltext{5}{Department of Physics, Carnegie Mellon University, 
5000 Forbes Ave., Pittsburgh, PA 15213.}

\altaffiltext{6}{Palomar Observatory, 105-24 Caltech, Pasadena, CA 91125.}

\begin{abstract}
We report preliminary results from an arcsecond-resolution X-ray survey of 
nearby galaxies using the Advanced CCD Imaging Spectrometer (ACIS)  on board 
the {\it Chandra X-ray Observatory.}\  The total sample consists of 41 
low-luminosity AGNs, including Seyferts, LINERs, and LINER/\hii\ transition 
objects.  In the initial subsample of 24 objects observed thus far, we detect 
in $\sim$62\% of the objects a compact, point-like source astrometrically 
coincident with either the optical or radio position of the nucleus.  The high 
detection rate strongly suggests that the majority of the objects do contain 
weakly active, AGN-like cores, presumably powered by central massive black 
holes.  The 2--10~keV luminosities of the nuclear sources range from 
$<10^{38}$ to 10$^{41}$ \lum, with a median value of 2\e{38} \lum.  Our 
detection limit corresponds to $L_{\rm X}$(2--10~keV) $\approx$ 8\e{37} \lum\ 
for the typical sample distance of 12~Mpc; this limit is two orders of 
magnitude fainter than the weakest sources of this kind previously studied 
using {\it ASCA}\ or {\it BeppoSAX}.  The new data extend toward lower 
luminosities the known linear correlation between hard X-ray and H\al\ 
luminosity for broad-line AGNs.  Many narrow-line objects do contain X-ray 
cores, consistent with either weak AGNs or X-ray binary systems, but they have 
X-ray luminosities a factor of 10 below the $L_{\rm X}$--$L_{{\rm H}\alpha}$ 
relation of the broad-line sources.  Their distributions of photon energies 
show no indication of exceptionally high absorption.  The optical line 
emission in these nuclei is likely powered, at least in part, by 
stellar processes.
\end{abstract}


\section{Introduction}
Knowledge of the local space density of active galactic nuclei (AGNs) impacts 
many astrophysical issues.  Surveys for nearby AGNs furnish critical data for 
quantifying the faint end of the AGN luminosity function, for characterizing 
the demography of massive black holes, and for investigating accretion 
physics.  Optical spectroscopic surveys indicate that many nearby galaxies 
possess mildly active nuclei (Ho, Filippenko, \& Sargent 1997b).  Although a 
significant fraction of these objects do contain accretion-powered sources 
qualitatively similar to more powerful AGNs such as classical Seyfert nuclei 
and quasars (Ho 1999), the physical nature of many still remains ambiguous.  
Observations at ultraviolet and optical wavelengths sometimes point to stellar 
processes as the underlying agent responsible for the activity (e.g., Maoz 
et al. 1998; Barth \& Shields 2000), but these data generally cannot rule out 
the presence of a highly obscured AGN component.  

Unless Compton-thick conditions prevail, hard X-ray data provide a much more 
definitive probe. In recent years, observations using {\it ASCA}\ and 
{\it BeppoSAX}\ have shed considerable light on the nature of low-luminosity 
AGNs (see Terashima 1999 and Ptak 2001 for reviews).  The fairly coarse 
angular resolution of these satellites, however, has biased detections to 
more luminous sources which are less contaminated by extended emission from 
the host galaxy.  Moreover, the relatively long integration times required for 
the observations necessarily restricted the published studies to small, 
usually X-ray selected, and possibly unrepresentative samples.

The exquisite imaging capability of the {\it Chandra X-ray Observatory}\ 
offers significant advantages for the study of low-luminosity AGNs.  
The good instrumental response for photon energies up to 8~keV permits
detection of faint AGN emission even if subject to absorption as high
as $N_{\rm H}\,\approx\,10^{24}$ cm$^{-2}$.  Thus, obscured AGNs can be
found.  Faint nuclei embedded deeply in the central regions of bright bulges 
can be detected reliably only under high angular resolution.  As we 
demonstrate in this Letter, {\it Chandra}\ can probe the nuclear regions of 
nearby galaxies effectively and efficiently, allowing us to survey, for the 
first time, a sizable sample of optically selected galaxies covering a wide 
range of nuclear activity.  With brief, ``snapshot'' exposures, we are able 
to detect or set stringent limits to compact nuclear X-ray sources with 
unprecedented sensitivity.  We revisit some longstanding, unresolved issues 
concerning the physical nature of low-luminosity AGNs in light of the new 
measurements.  The sharp X-ray images additionally reveal a plethora of 
previously unknown structural details in the circumnuclear environment of 
nearby galaxies.

 
\vspace*{0.3cm}

\begin{figure*}[t]
\centerline{\psfig{file=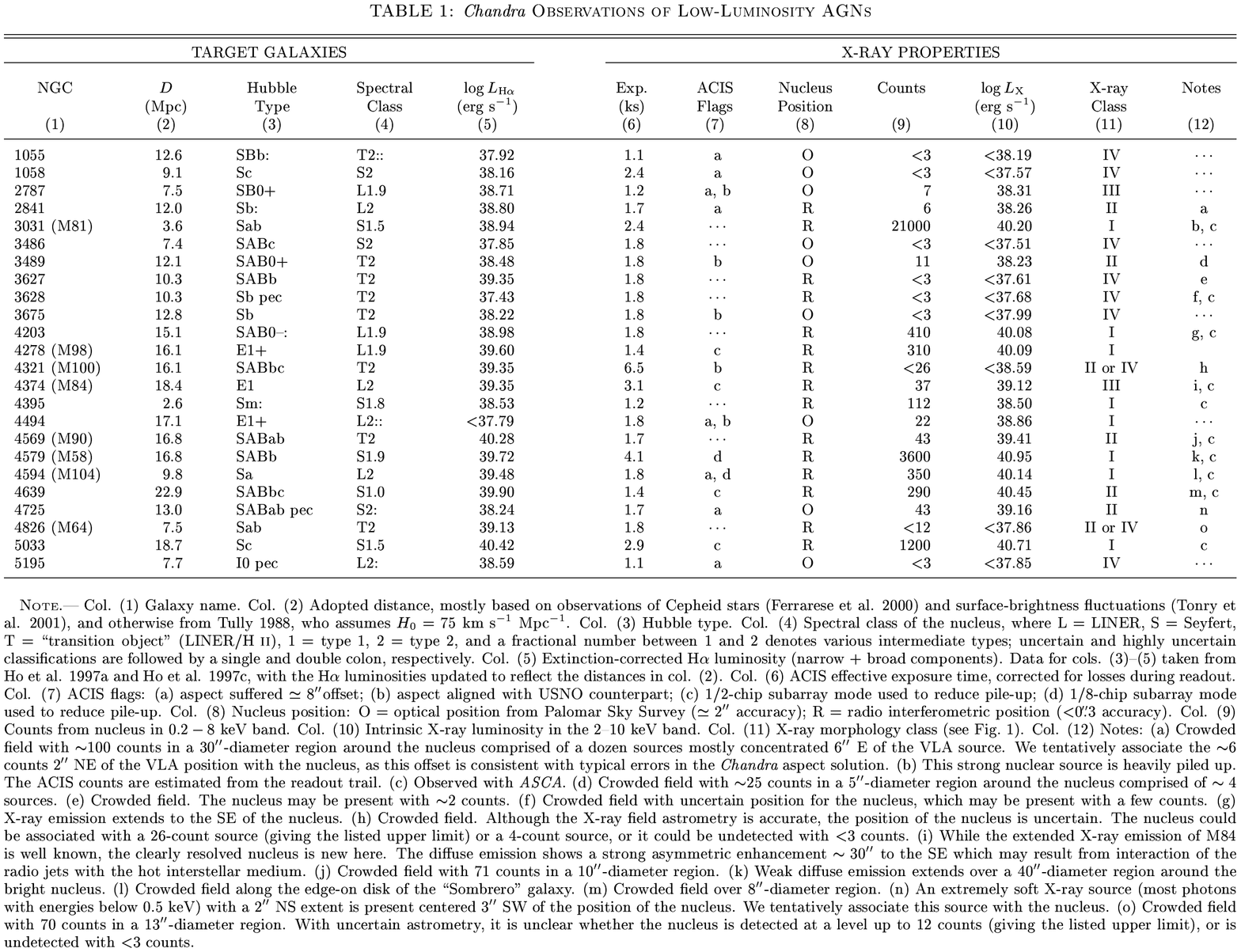,width=19.5cm,angle=90}}
\end{figure*}
\vspace*{0.3cm}
 

\section{Observations and Data Analysis}

Our targets were selected from the Palomar survey of nearby galaxies, a
sensitive spectroscopic study of a nearly complete sample of 486 bright
($B_T\,\leq$ 12.5 mag), northern ($\delta\,>$ 0\deg) galaxies (Ho, Filippenko,
\& Sargent 1997a,b).  As part of the Guaranteed Observer program for the 
Advanced CCD Imaging Spectrometer (ACIS; Garmire et al. 2001)
team, we chose a sample of 41 emission-line nuclei considered to be AGN
candidates, including Seyfert nuclei, low-ionization nuclear emission-line 
regions (LINERs; Heckman 1980), and LINER/\hii\ ``transition'' nuclei (Ho et 
al. 1997b).  Thirty-five objects belong to a complete, volume-limited sample 
within 13 Mpc; the remaining were included as prototypes of different 
classes.  To date, observations have been acquired for 24 galaxies.  While 
this subsample is at the moment somewhat heterogeneous, it does form a 
representative subset of nearby objects with weak nuclear activity, and it 
covers a mixture of Hubble types.  The sample, summarized in Table~1, is 
equally divided among Seyferts, LINERs, and transition objects; of these, 
1/3 are ``type~1'' and 2/3 are ``type~2'' sources, nuclei with and without 
evidence for broad emission lines, respectively.  The median distance is 12 Mpc.

Each galaxy was observed with ACIS on board {\it Chandra}\ (Weisskopf, O'Dell, 
\& van~Speybroeck 1996).  We used the on-axis backside-illuminated CCD chip
in the spectroscopic array because it is more sensitive to soft 
(0.2~keV $<\,E\,<$ 0.5~keV) X-rays than the frontside-illuminated chips in the
imaging array.  Most observations used the standard mode of reading out the 
full chip every 3.2~s, but in six cases where a bright nuclear source was 
known from {\it ROSAT}\ or {\it ASCA}\ studies, only a portion of the chip was 
read out more frequently in order to reduce photon pile-up.  Exposures of 
2~ks were requested, but the actual integrations ranged from 1.1 to 6.5~ks 
due to constraints on satellite scheduling. 

Our treatment of the data starts with the Level~1 event file produced by the 
{\it Chandra}\ X-ray Center.  We first apply a correction for charge-transfer 
inefficiency which reduces biases in event energies and grades 
(Townsley et al. 2000).  Next, the data are cleaned of events with 
bad status flags, hot columns, and bad flight grades from cosmic-ray
impacts.  We then select {\it ASCA}\ grades 02346 and energies 0.2--8~keV for
science analysis.  A search for sources is conducted in both the target
and ancillary CCD chips using a wavelet-transform algorithm for source 
detection (Freeman et al. 2001).  Unresolved on-axis sources can be

%
%
%

 

\psfig{file=fig1_new.ps2,width=9.5cm,angle=0}
\figcaption[fig1_new.ps2]{Representative ACIS images of the inner regions of
nearby galaxies hosting low-luminosity AGNs, chosen to exemplify the four X-ray
classes: I = dominant nucleus; II = nucleus comparable in brightness to
off-nuclear sources in the galaxy; III = nucleus embedded in diffuse
emission; and IV = nucleus absent.  Each panel subtends $90\arcsec \times
90\arcsec$, and the 4\asec-diameter circle is centered on the radio or
optical position of the nucleus.
\label{fig1}}
\vskip 0.3cm


\noindent
reliably located as faint as $\simeq 4$ counts.  Nuclear counts were typically
extracted from a 2\arcsec-diameter circle without background subtraction, but
certain cases required special treatment (see notes to Table~1).  ACIS count
rates are converted to X-ray luminosities assuming an intrinsic power-law
spectrum with photon index $\Gamma = 1.8$ and absorption from our Galactic
interstellar medium with column density $N_{\rm H} = 2 \times 10^{20}$
cm$^{-2}$. These assumptions\footnote{The hard X-ray spectra of most
low-luminosity AGNs are well fit by $\Gamma\,\approx\, 1.8$, and in many cases
no significant absorption in excess of the Galactic foreground is indicated
(e.g., Terashima 1999). None of our objects suffer heavy Galactic extinction;
the median $N_{\rm H} \approx 2 \times 10^{20}$ cm$^{-2}$.  The luminosities
would increase by a factor of 1.8 and 6.7 for intrinsic columns of 2\e{21}
and 2\e{22} cm$^{-2}$, respectively.} give $L_{\rm X}$(2--10~keV) = $3.6
\times 10^{37}$ erg s$^{-1}$ (ACIS~cts/ks) ($D$/10~Mpc)$^2$.  We quote X-ray
luminosities in the 2--10~keV range for ease of comparison with literature data.

Examination of the images quickly revealed that, in many cases, the
nucleus is only one, and not necessarily the brightest, of many X-ray
sources or structures in the core regions of the target galaxies.
Astrometric alignment of the X-ray image to the galaxy is thus
critical.  The satellite aspect system based on star-tracker cameras
gives absolute astrometric alignments in the {\it Hipparcos}\ frame with
accuracies usually within $\pm 2$\arcsec.  But in a few cases, X-ray
sources unrelated to the galaxy are found to be associated with
foreground stars ($R \simeq 10-15$ mag) with positions from the USNO A-2
catalog accurate to $\pm$0\farcs3.  For most of the target galaxies, the 
location of the nucleus is known to high accuracy ($\pm$0\farcs3 and 
often $<$0\farcs1) from radio observations.  But for some, no radio source 
is found, and positions are based on relatively inaccurate measurements
from sky survey Schmidt plates (around $\pm 2$\arcsec).  Thus, our ability to 
align the ACIS image to the galaxy ranges from a few tenths to several 
arcseconds.  A more detailed discussion of the X-ray analysis will be 
presented for the full volume-limited sample in a forthcoming paper.



\psfig{file=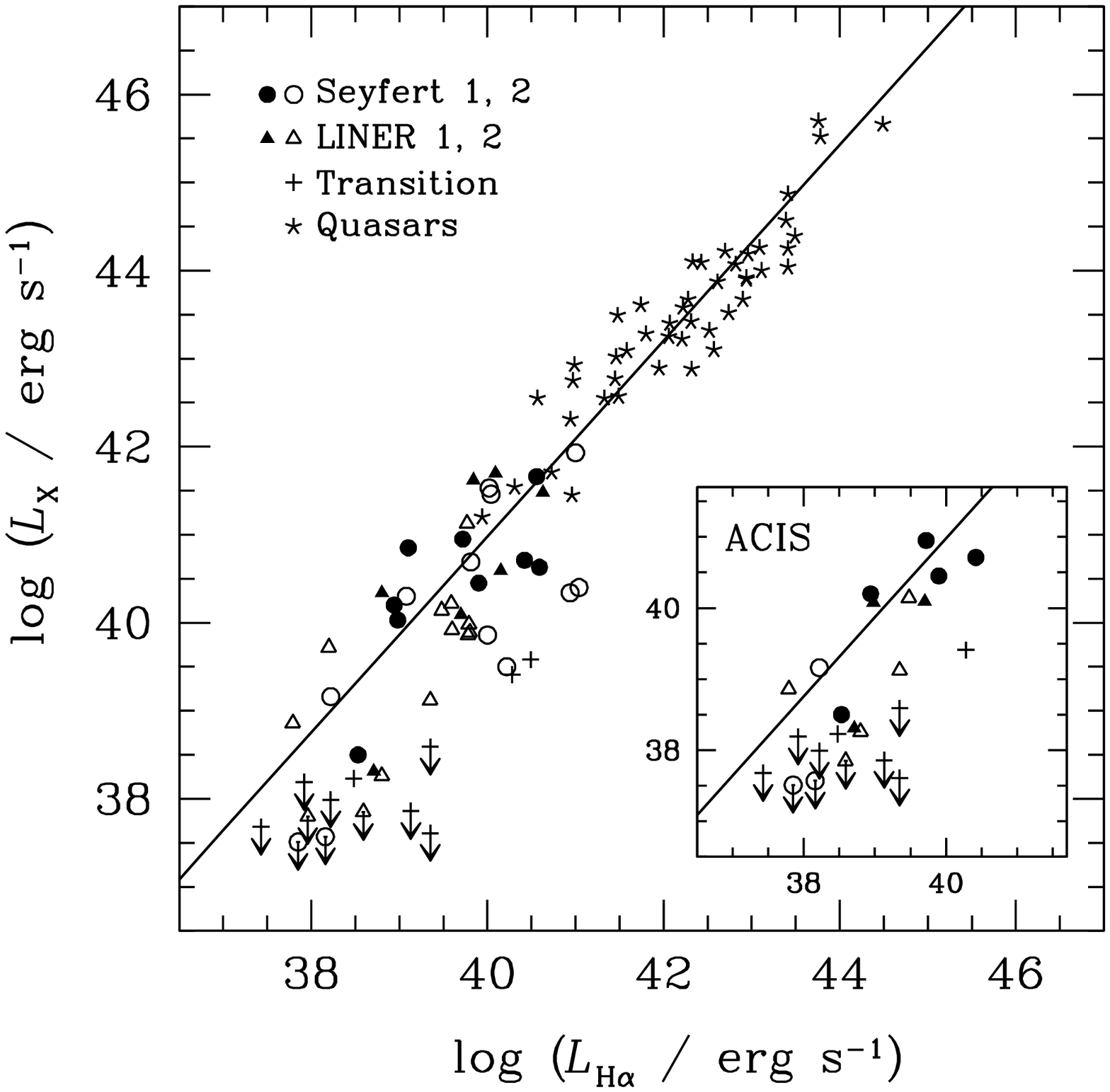,width=8.5cm,angle=0}
\figcaption[fig2_new.ps]{
The $L_{\rm X}$--$L_{{\rm H}\alpha}$ correlation for AGNs.
The X-ray luminosity represents the intrinsic (unabsorbed) power-law component
in the 2--10~keV band.  The H\al\ luminosity includes both the narrow and
broad (if present) components of the line, corrected for extinction due to
the Galaxy and the narrow-line region.  The data for low-$z$ quasars are taken
from Ward et al. (1988), adjusted to $H_0$ = 75 \kms\ Mpc$^{-1}$.  The
low-luminosity sources (low-luminosity Seyferts, LINERs, and transition
nuclei) come from this paper and Terashima et al. (2000a), whose distances
have been updated according to the precepts outlined in Table~1.   The
{\it solid line}\ shows the best-fit unweighted linear regression line (see
text) for the type~1 objects and low-$z$ quasars.  The insert on the lower
right corner of the graph highlights the new ACIS observations for clarity.
\label{fig2}}
\vskip 0.3cm


\section{Results and Discussion}

The morphologies of the X-ray images can be loosely grouped into four
classes (Fig.~1): (I) dominant nuclear source; (II) nuclear source comparable
in brightness to off-nuclear sources in the galaxy; (III) nuclear source
embedded in diffuse emission; and (IV) no nuclear source.  A compact X-ray
source can be associated with the optical or radio nucleus in 62\% 
(15/24) of the galaxies.  It is of interest to compare the detection frequency 
as a function of AGN spectral type.  All eight of the broad-line nuclei (3 
LINER 1s, 5 Seyfert 1s) were detected, by contrast to only 44\% (7/16) of the 
narrow-line sources (1/3 Seyfert 2s, 4/5 LINER 2s, 2/8 transition objects).  
For LINERs as a class, the detection rate ranges from $\sim$60\% to 
$\sim$90\%, depending on whether we include or exclude transition objects as 
LINERs.  

The gross properties of the optical emission-line spectra of AGNs can be 
explained by photoionization from the central AGN continuum.  In support of 
this picture, the strength of the hydrogen recombination lines generally 
scales with the X-ray luminosity in powerful Seyfert 1 nuclei and quasars 
(e.g., Kriss, Canizares, \& Ricker 1980; Ward et al. 1988).  The 
$L_{\rm X}$--$L_{{\rm H}\alpha}$ correlation has been shown to extend down to 
the regime of low-luminosity AGNs, both in the soft X-ray band (Koratkar et 
al. 1995; Roberts \& Warwick 2000; Halderson et al. 2001) and in the hard 
X-ray band (Terashima, Ho, \& Ptak 2000a)\footnote{As these papers show, the 
$L_{\rm X}$--$L_{{\rm H}\alpha}$ correlation is not an artifact of distance 
effects.}.  This is consistent with the idea that low-luminosity AGNs share the 
same basic physical processes as in high-luminosity AGNs.  Terashima et al. 
(2000a,b) find, however, that type~2 sources generally show systematically
lower values of $L_{\rm X}/L_{{\rm H}\alpha}$ compared to type~1 objects.  

Our sample reinforces these conclusions, as shown in Figure~2.  Low-luminosity 
Seyfert~1s and LINER~1s trace the $L_{\rm X}$--$L_{{\rm H}\alpha}$ relation 
to $\log L_{\rm X}\,\approx$ 38.5, with a median $L_{\rm X}/L_{{\rm H}\alpha}$ 
of 15; the slope is close to, but slightly steeper than, unity.  The best-fit 
unweighted linear regression line, calculated using the ordinary least-squares 
solution bisector with jackknife resampling (Feigelson \& Babu 1992), for 
type~1 objects (including low-$z$ quasars) is
$\log L_{\rm X}\,=\,(1.11\pm0.054) \log L_{{\rm H}\alpha}\,-\,(3.50\pm2.27)$.
The detected type~2 objects loosely follow the same correlation, but with 
somewhat greater scatter and offset toward lower $L_{\rm X}/L_{{\rm H}\alpha}$ 
(median $\sim$2).  The ACIS observations contribute nine stringent upper 
limits in the regime $\log L_{\rm X}$ \lax 38.0, all of which deviate 
significantly from the correlation established by the type~1 sources (most 
have $L_{\rm X}/L_{{\rm H}\alpha}\,<\,1$).  The majority of the upper limits 
are transition objects. 

Note that the intrinsic scatter in the $L_{\rm X}$--$L_{{\rm H}\alpha}$ 
relation should be significantly less than indicated in Figure~2.  The 
photometric errors associated with the luminosities for any individual object 
may be substantial (see footnote 7 for $L_{\rm X}$ and Ho et al. 1997a for 
$L_{{\rm H}\alpha}$).  The non-simultaneity of the X-ray and optical 
observations likely introduces additional scatter.  Although the variability 
characteristics of low-luminosity AGNs are poorly constrained, some of them 
do vary, at least in the X-rays (e.g., Ptak et al. 1998).  The H\al\ 
measurements of Ho et al. (1997a) were extracted from a 2\asec$\times$4\asec\ 
aperture, somewhat larger than that used for the X-rays, but the effect of 
aperture mismatch should be minimal because the optical line emission 
tends to be highly centrally concentrated (Pogge et al. 2000). 

The present sample is still small and incomplete, and we caution against 
premature generalizations based on the above statistics.  Nonetheless, these 
early results are in broad agreement with the following:

(1) Most, perhaps all, low-luminosity type~1 objects, both Seyferts and 
LINERs, are genuine AGNs similar to classical Seyfert~1 nuclei and quasars.

(2) A significant fraction of low-luminosity type~2 sources do contain 
a central X-ray core, consistent with the presence of an AGN, but they 
are underluminous in X-rays compared to type~1 nuclei of the same H\al\ 
luminosity.  This suggests that the optical line emission may not be 
powered exclusively by a central AGN. Although detailed spectral fitting was 
not performed, examination of the photon energies indicates that few, if any, 
of the nuclear sources are absorbed by columns in the range 
$N_{\rm H}\,\approx\, 10^{22}-10^{24}$ cm$^{-2}$.  Our data thus tentatively 
suggest that the standard unified model for Seyferts (Antonucci 1993)
may not hold at very low luminosities.  But we cannot exclude the presence 
of spectral components with $N_{\rm H}$ \gax $10^{25}$ cm$^{-2}$, nor the 
possibility that the X-ray emission in these systems arises entirely from 
X-ray binaries without any AGN component.

(3) Most transition objects are unrelated to AGNs.

(4) At least 60\% of LINERs contain AGNs, consistent with the estimates of 
Ho (1996, 1999).

\acknowledgements
This work was sponsored by NASA contract NAS 8-38252 (Garmire, PI).
The research of L.~C.~H. and A.~V.~F. is partly funded by NASA LTSA grant 
NAG 5-3556, and by NASA grants GO-06837.01-95A, AR-07527.02-96A, and 
AR-08361.02-97A from the Space Telescope Science Institute (operated by AURA, 
Inc., under NASA contract NAS5-26555).  W.~N.~B. acknowledges support from 
NASA LTSA grant NAG 5-8107.
%

%

%
%

\end{document}